\documentclass[journal,comsoc]{IEEEtran}
\usepackage{cite}
\usepackage{graphicx}
\usepackage{amsmath}
\usepackage{times}
\usepackage{latexsym}
\usepackage{graphicx}
\usepackage{bm}
\usepackage{amssymb}
\usepackage[center]{caption2}
\usepackage{stfloats}
\usepackage{cases}
\usepackage{subfigure}
\usepackage{array}
\usepackage{setspace}
\usepackage{fancyhdr}
\usepackage{citesort}
\usepackage{colortbl}
\usepackage{CJK}
\usepackage{tikz}
\usepackage{float}
\usepackage{color}
\usepackage{bm}
\usepackage{mathrsfs}
\usepackage{diagbox}

\renewcommand{\QED}{\QEDopen}

\allowdisplaybreaks[4]

\renewcommand{\baselinestretch}{0.97}

\newcaptionstyle{mystyle2}{%
\captionlabel $.$ \, \captiontext \par} \captionstyle{mystyle2}
\renewcommand\thesubsection{\Alph{subsection}}

\begin{document}
\title{Unsupervised Learning for Passive Beamforming}
\author{\authorblockN{Jiabao Gao, Caijun Zhong, Xiaoming Chen, Hai Lin and Zhaoyang Zhang}}
\maketitle

\begin{abstract}
Reconfigurable intelligent surface (RIS) has recently emerged as a promising candidate to improve the energy and spectral efficiency of wireless communication systems. However, the unit modulus constraint on the phase shift of reflecting elements makes the design of optimal passive beamforming solution a challenging issue. The conventional approach is to find a suboptimal solution using the semi-definite relaxation (SDR) technique, yet the resultant suboptimal iterative algorithm usually incurs high complexity, hence is not amenable for real-time implementation. Motivated by this, we propose a deep learning approach for passive beamforming design in RIS-assisted systems. In particular, a customized deep neural network is trained offline using the unsupervised learning mechanism, which is able to make real-time prediction when deployed online. Simulation results show that the proposed approach maintains most of the performance while significantly reduces computation complexity when compared with SDR-based approach.
\end{abstract}


\section{INTRODUCTION}
With the commercialization of fifth generation wireless communication systems, how to reduce the network deployment cost and energy consumption has stood out as one of the major challenges for future sustainable and green wireless systems. Recently, the reconfigurable intelligent surface (RIS) has emerged as a promising candidate to tackle the above challenges thanks to low manufacturing, hardware and energy cost.\footnote{Please note, there have been other terminologies for RIS, such as intelligent reflecting surface \cite{baseline}, intelligent wall \cite{wall}, passive intelligent mirror \cite{mirror}, and reconfigurable metasurface \cite{metasurface}.} Specifically, the RIS is a planar array composed of a large number of low-cost reconfigurable reflecting elements, which are able to modify the phase shift of the incident signal. Through proper adjustment of the phase shifts, the reflected signal from the RIS can add coherently with the signal from the direct path at the intended user to substantially improve the receive signal strength.

It is worth noting that RIS is a brand new technology that significantly differs from other related technologies such as amplify-and-forward (AF) relaying and backscatter communications. For instance, compared to AF relaying, the RIS does not need to generate its own transmission signal, but passively reflects the incident signal, hence having much lower power consumption. Compared to the backscatter communications, the RIS does not deliver any of its own information, but only acts as a helper to enhance the performance of existing links \cite{baseline}.

Apparently, the design of phase shifts, also known as passive beamforming, is of critical importance for RIS-assisted communication systems. However, the non-convexity introduced by unit modulus constraint makes the derivation of optimal solution difficult. In \cite{baseline}, a suboptimal solution for passive beamforming in RIS-assisted single user multiple-input single-output downlink systems is proposed using the conventional semi-definite relaxation (SDR) technique. However, the SDR-based approach is computationally expensive and is not amenable for real-time implementation. In \cite{mirror}, the authors investigated the multi-user scenario and addressed the problem of maximizing the achievable rate by combining alternating maximization with the majorization-minimization method. Later on, by imposing a rank-1 constraint on the channel between the source and RIS, a closed-form analytical solution was derived in \cite{baseline2}. Besides, energy efficiency for RIS-assisted systems was investigated in \cite{EE}, while \cite{powerminimization} tackled the problem of minimizing the total transmit power subject to individual signal-to-interference-plus-noise ratio constraint. In practice, discrete phase shifts are used due to hardware constraint. In this regard, \cite{discrete1} and \cite{discrete2} have investigated performance of RIS systems with only a finite number of phase shift levels.

In the past few years, deep learning (DL) has demonstrated its remarkable potential in dealing with non-convex problems \cite{BFNN,R1}. In addition, DL based approach enables fast computation compared with the traditional iterative algorithms \cite{speed}. These desirable features make it appealing for many practical applications in wireless communications\cite{WC2}. In the context of RIS-assisted wireless communications, a supervised learning based approach was proposed in \cite{SupervisedBF} where a deep neural network (DNN) is trained offline to establish the implicit relationship between the measured coordinate information and RIS's phase configuration. Nevertheless, a major issue for supervised learning is how to obtain labels. In \cite{SupervisedBF}, the optimal labels were obtained via exhaustive search, which is extremely expensive to implement in practice, especially if large number of training samples are required.

To avoid the labelling overhead of supervised learning, in this paper, we propose to adopt the unsupervised learning mechanism\cite{BFNN,UNPA} for passive beamforming design, where no labels are required. In particular, a customized DNN architecture is developed for the passive beamforming design problem, and a set of tailored features are selected for the training process. Simulation results show that the proposed unsupervised learning based approach requires much less computational time with tolerable performance deterioration when compared with the conventional SDR-based approach.

The rest of this paper is organized as follows. Section II introduces the RIS-assisted wireless communication system model and the problem formulation of passive beamforming. In Section III, conventional approaches for the passive beamforming problem in single antenna case and multi-antenna case are briefly introduced. The unsupervised learning based approach is proposed in Section IV, and Section V presents simulation results to evaluate the performance of the proposed approach. Finally, the paper is concluded in Section VI.

\emph{Notations:} Scalars, vectors and matrices are denoted by italic letters, bold-face lower-case and bold-face upper-case letters, respectively. $\mathbb{C}^{x \times y}$ denotes the space of $x \times y$ complex vectors or matrices. $||\cdot||$ denotes the Euclidean norm and $\text{diag}\left(\cdot\right)$ denotes the diagonalization of a vector. Superscript $*$, $T$ and $H$ denote conjugate, transpose and conjugate transpose respectively. $\text{tr}\left(\cdot\right)$ denotes the trace of a matrix and $\bm{X}\succeq 0$ means that $\bm{X}$ is positive-semidefinite. $\mbox{Norm}\left(\cdot\right)$ is the operation of normalizing a complex scalar's modulus, and $\bm{x}[1:N]$ fetches the first $N$ elements of vector $\bm{x}$. $\mathcal{CN}(\mu,\sigma^2)$ denotes the distribution of a circularly symmetric complex Gaussian random variable with mean $\mu$ and covariance $\sigma^2$.

\section{System model and Problem formulation}
\label{System model and problem formulation}
\begin{figure}[htbp]
\centering
\includegraphics[width=0.450\textwidth]{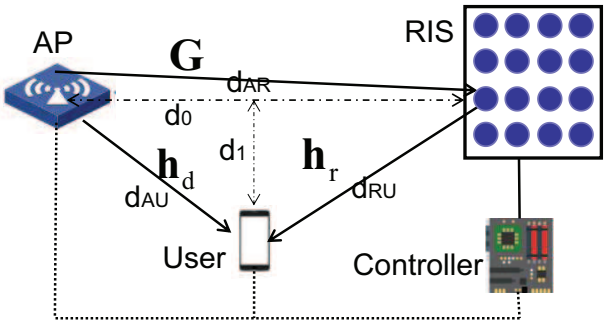}
\caption{System model.}
\label{SystemModel}
\end{figure}

We consider a three-node system consisting of an access point (AP) equipped with $M$ antennas, a RIS equipped with $N$ reflecting elements and a single antenna user as illustrated in Fig. \ref{SystemModel}. A controller connecting the AP and RIS is used to adaptively adjust the phase shifts of reflecting elements and coordinate the switching between the receiving mode for channel estimation and the reflecting mode for signal reflection \cite{wall}. The baseband receive signal at the user is the superposition of the direct signal from the AP and the reflected signal from RIS, which can be expressed as

\begin{equation}
y=(\bm{G\Theta}\bm{h}_r+\bm{h}_d)^T\bm{w}s+n ,
\end{equation}
where $\bm{w} \in \mathbb{C}^{M \times 1}$ denotes the transmit beamforming vector satisfying $\left\|\bm{w}\right\|^2 \leq p$, and $s$ is the information symbol with unit power drawn from a certain constellation set. Also, $\bm{h}_d \in \mathbb{C}^{M \times 1}$, $\bm{h}_r \in \mathbb{C}^{N \times 1}$ and $\bm{G} \in \mathbb{C}^{M \times N}$ denote the channels of the direct link between AP and user, the reflecting link between RIS and user, and the link between RIS and AP, respectively. All the channels are assumed to be quasi-static and flat-fading. In practice, the channel state information can be obtained by different methods as pointed out in \cite{CSI}. Moreover, $\bm{\Theta}=\text{diag}\{\bm{\theta}\}$ is the phase shift matrix of the RIS, where $\bm{\theta}=[\theta_1,\theta_2, \ldots, \theta_N]^T \in \mathbb{C}^{N \times 1}$ and $|\theta_n|=1$. The additive noise $n \sim \mathcal{CN}(0,\sigma^2)$. The signals reflected by the RIS for two or more times are ignored due to severe path loss \cite{baseline}. Therefore, the receive SNR $\gamma$ at the user can be expressed as
\begin{equation}
\gamma = \frac{1}{\sigma^2}|(\bm{G\Theta}\bm{h}_r+\bm{h}_d)^T\bm{w}|^2.
\label{SNR}
\end{equation}

Given $\Theta$, it is well known that maximum-ratio transmission (MRT) is the optimal transmit beamforming strategy \cite{mrt}, i.e., ${\bm{w}_\text{opt}}^T=\sqrt{p}\frac{(\bm{G\Theta}\bm{h}_r+\bm{h}_d)^H}{\left\|\bm{G\Theta}\bm{h}_r+\bm{h}_d\right\|}$. Substituting $\bm{w}_{\text{opt}}$ into Equation (\ref{SNR}), the optimal phase shift $\bm{\theta}$ at the RIS maximizing the transmission rate is the solution of the following optimization problem
\begin{align}
\text{P1}: \qquad
& \underset{\bm{\theta}}{\text{max}}
& & \left\|\bm{G\Theta}\bm{h}_r+\bm{h}_d\right\|^2 \label{snr_final}\\
&\ \, \text{s.t.}
& & \, |\theta_n|=1, \; n = 1, \ldots, N.
\end{align}

\section{Conventional Approach}
Problem P1 is a typical non-convex quadratically constrained quadratic program (QCQP), hence the optimal solution is intractable except for the single antenna case. For $M\geq 2$, the common approach is to find a suboptimal solution using techniques such as SDR.

\subsection{Single antenna case}
\label{SingleAntennaCase}
When $M=1$, problem P1 can be rewritten as
\begin{align}
\text{P2}: \qquad
& \underset{\bm{\theta}}{\text{max}}
& & |r|^2 \\
&\ \, \text{s.t.}
& & |\theta_n|=1, \; n = 1, \ldots, N,
\end{align}
where $r\triangleq\sum\limits_{n=1}^N g_nh_{r_n}\theta_n + h_d$. It is easy to show that the maximum can be achieved by aligning the phases of ${g}_n {h}_{r_n}{\theta}_n$ to that of $h_d$. Hence, the optimal phase shift ${\theta}_n$  can be computed as
\begin{equation}
{\theta}_n^{opt}=\text{Norm}\left(\frac{h_d}{{g}_n {h}_{r_n}}\right),
\label{OptimalSolution}
\end{equation}

\subsection{Multi-antenna case}
We now consider the multi-antenna case, i.e., $M \ge 2$. According to \cite{baseline}, problem P1 can be reformulated as the following homogeneous QCQP
\begin{align}
\text{P3}: \qquad
& \underset{\bm{\bar{\theta}}}{\text{max}}
& & \bm{\bar{\theta}}^H \bm{R} \bm{\bar{\theta}} \\
&\ \, \text{s.t.}
& & |\theta_n|=1, \; n = 1, \ldots, N,
\end{align}
where \begin{gather*}\bm{\bar{\theta}}=\begin{bmatrix} \bm{\theta} \\ t \end{bmatrix}, \bm{R}=\begin{bmatrix} \bm{D}_h^H\bm{G}^H\bm{G}\bm{D}_h & \bm{D}_h^H\bm{G}^H\bm{h}_d\\ \bm{h}_d^H\bm{G}\bm{D}_h & 0\end{bmatrix},\end{gather*}
$\bm{D}_h=\text{diag}\{{\bm{h}_r}\}$, and $t$ is an auxiliary variable.

Define $\bm{Q} \triangleq\ \bm{\bar{\theta}}\bm{\bar{\theta}}^H$, we have $\bm{\bar{\theta}}^H \bm{R} \bm{\bar{\theta}} = \text{tr}(\bm{RQ})$. $\bm{Q}$ is a positive-semidefinite matrix with rank one. By relaxing the rank-one constraint, P3 can be converted to the following convex problem
\begin{align}
\text{P4}: \qquad
& \underset{\bm{Q}}{\text{max}}
& & \text{tr}(\bm{RQ}) \\
&\ \, \text{s.t.}
& & \bm{Q} \succeq 0; \bm{Q}_{n,n}=1, \; n = 1, \ldots, N+1.
\end{align}

From the optimal solution of P4, a suboptimal solution of P3 can be obtained through the technique of randomization. Then, the suboptimal solution of of P1 can be obtained as $\bm{\theta}=\text{Norm}(\bm{\bar{\theta}}[1:N]/\bm{\bar{\theta}}_{N+1})$.

\section{Unsupervised learning based approach}
\label{system}
The performance of suboptimal solutions obtained by conventional optimization based approaches can not be guaranteed, and they in general incur high complexity. Motivated by these issues, we propose a DL based framework to tackle problem P1.

\subsection{Feature Design}
From Equation (\ref{snr_final}), it is intuitive to use the channels $\bm{G}$, $\bm{h}_r$ and $\bm{h}_d$ as the input. However, such a simple approach turns out to be ineffective and problematic.

To see this, let us define $\bm{r} \triangleq\ \bm{G\Theta}\bm{h}_r+\bm{h}_d$, then the $i$-th element of $\bm{r}$ can be written as
\begin{equation}
r_i = \sum\limits_{n=1}^Ng_{i,n}h_{r_n}\theta_n + h_{d_i}, \; i = 1, \ldots, M,
\label{obj}
\end{equation}
which explicitly shows the product structure of $\bm{G}$ and $\bm{h}_r$. Hence, instead of simply choosing $\bm{G}$, $\bm{h}_r$, a more appropriate feature is to use the product of the elements of $\bm{G}$ and $\bm{h}_r$. In addition, the real and imaginary parts are treated as separate features. As such, the final feature vector is denoted by $\bm{F} \in \mathbb{C}^{2(NM+M) \times 1}$. It is worth pointing out that the above feature design not only reduces the dimension of input, but also exploits the inherent structural information, therefore substantially improves the training efficiency and network performance.

\subsection{Loss Function}
The loss function is defined as
\begin{equation}
\text{Loss} = - \frac{1}{K}\sum\limits_{k=1}^K \left\|\bm{G}^k\bm{\Theta}^k\bm{h}_r^k+\bm{h}_d^k\right\|^2,
\end{equation}
where $K$ is the number of training samples in a mini batch. Please note, unlike the commonly used mean square error in supervised learning systems, the loss function is set to be the negative of the objective function in P1, which reflects the unsupervised nature of the proposed approach. Besides, the following Lambda layer is implemented to convert the predicted real phase shift vector $\bm{p}_{pred}$ to its complex form $\bm{\theta}_{pred}$ for loss computation\cite{BFNN}
\begin{equation}
\label{lambda}
\bm{\theta}_{pred} = e^{j \cdot \bm{p}_{pred}} = \cos(\bm{p}_{pred}) + j \cdot \sin({\bm{p}_{pred}}).
\end{equation}

\subsection{Network Architecture and training}
\begin{figure*}[htbp]
\centering
\includegraphics[width=0.85\textwidth]{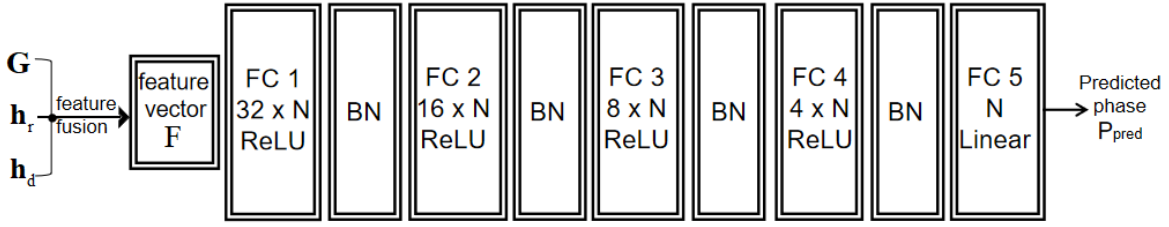}
\caption{Network architecture.}
\label{Network}
\end{figure*}

Fig. \ref{Network} illustrates the adopted architecture of the neural network, which is termed as ``RISBFNN". In particular, RISBFNN is made up of 5 fully-connected (FC) layers with $32N$, $16N$, $8N$, $4N$ and $N$ neurons respectively. The idea of setting the number of neurons being proportional to $N$ is to ensure adequate learning ability when the system scales. For activation function, the first four FC layers adopt the rectified linear unit (ReLu), while FC5 uses a Linear unit to output the phase shift prediction.

To train the network, adam optimizer with initial learning rate 0.001 is used. Also, the number of maximal epoch is set to 1000, and early stopping with patience 30 is applied to improve the training efficiency. In addition, to expedite the convergence, the learning rate decays by a factor of 0.33 whenever the validation loss does not decrease for a consecutive 15 epochs.

\begin{figure}[htbp]
\centering
\includegraphics[width=0.45\textwidth]{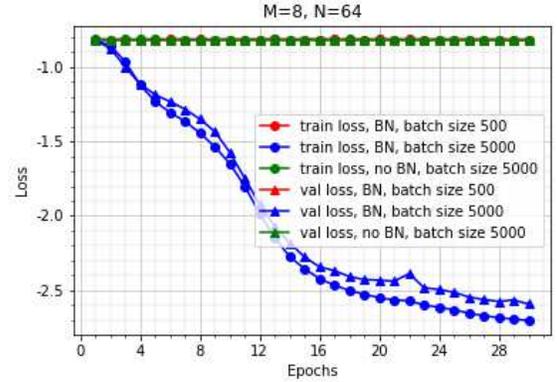}
\caption{Impact of BN and batch size on loss.}
\label{loss}
\end{figure}

During network training, it turns out that BatchNormalization (BN) layer\cite{BN} and batch size are two key hyperparameters to make RISBFNN work effectively. An exemplary training process is illustrated in Fig. \ref{loss} where $M=8, N=64$. As can be readily observed, without BN layer and sufficiently large batch size, both the training and validation loss cannot decrease. Through extensive experiments, we found that a BN layer after each FC layer and a batch size of 5000 work well under various settings.

\begin{figure}[htbp]
\centering
\includegraphics[width=0.45\textwidth]{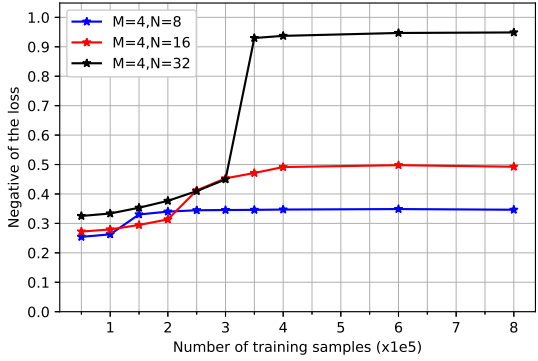}
\caption{Impact of the number of training samples.}
\label{training}
\end{figure}

The impact of the number of training samples is illustrated in Fig. \ref{training}. As can be observed, the performance settles when the number of training samples is sufficiently large. If the number of training samples is small, then overfitting occurs which substantially degrades the system performance. In addition, the minimum required sample number is configuration dependent, for instance, when $M=4, N=8$, 150000 samples are sufficient, while for the case $M=4, N=32$, 350000 samples are required. It is also worth highlighting that, since unsupervised learning is adopted, no labels are required, which significantly reduces the cost of obtaining training samples.

Taking training efficiency, testing performance and stability into consideration, we generate 800000, 200000 and 10000 samples for training, validation and testing respectively. It is also worth mentioning that, prior to training, standardization preprocessing is performed on each dimension of features by subtracting its average and dividing its standard deviation. All the training is executed by a GeForce GTX 1080 Ti GPU.

\section{Simulation Results}
\label{simulation}
In this section, simulation results are provided to demonstrate the performance of the proposed unsupervised learning approach.\footnote{For reproducible research, all source codes can be found at https://github.com/EricGJB/UN-Based-RISBF} We consider a similar indoor scenario as \cite{SupervisedBF}, where all the channels are modeled by independent Rayleigh small-scale fading, and the path loss in dB is computed as $20.4\log_{10}(d/d_{\sf ref})$ \cite{channelmodel}, with $d$ being the distance between transmitter and receiver in meters and $d_{\sf ref}= 1\mbox{m}$ denoting the reference distance. As illustrated in Fig .\ref{SystemModel}, the distance from AP to RIS is denoted by $d_\text{AR}$, while the distance from AP to user and from RIS to user can be computed as $d_\text{AU}=\sqrt{d_0^2+d_1^2}$ and $d_\text{RU}=\sqrt{(d_{\text{AR}}-d_0)^2+d_1^2}$ respectively, where $d_1$ is the vertical distance from user to the horizontal connection line of AP and RIS while $d_0$ is the distance from AP to the intersection. For the simulations, $d_{\text{AR}}$ is set to be $8\text{m}$, while $d_0$ and $d_1$ follows uniform distribution with range $[0, 8]$ and $[1, 6]$ respectively. Besides, $p/\sigma^2 = 10$ dB.

\subsection{Impact of $N$}
\begin{figure}[htbp]
\centering
\includegraphics[width=0.450\textwidth]{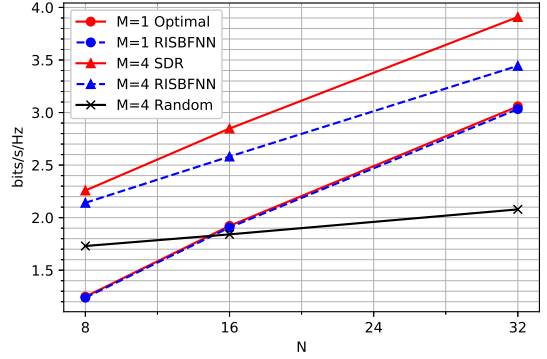}
\caption{Impact of the number of reflecting elements $N$.}
\label{VaryN}
\end{figure}

Fig. \ref{VaryN} illustrates the impact of the number of reflecting elements $N$ on the performance. As can be readily observed, the performance of both the RISBFNN and SDR approaches improve as $N$ becomes larger, which is intuitive since increasing $N$ can enhance the effective gain of the reflecting path. For the single antenna case, the two curves almost overlaps, which indicates that the proposed RISBFNN can achieve near optimal performance. For the multi-antenna case, the performance gap between SDR and RISBFNN becomes more significant as $N$ increases.

\subsection{Impact of $M$}
\begin{figure}[htbp]
\centering
\includegraphics[width=0.45\textwidth]{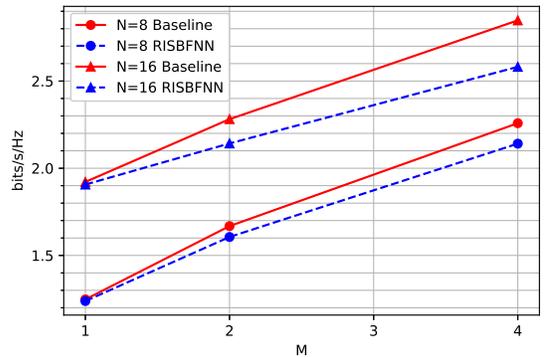}
\caption{Impact of antenna number $M$ on the performance. Blue and red curves represent the RISBFNN and baseline methods (optimal solution in Equation (\ref{OptimalSolution}) when $M=1$ and SDR approach when $M \ge 2$) respectively.}
\label{VaryM}
\end{figure}

Fig. \ref{VaryM} depicts the impact of $M$ on the performance. As expected, we see that the performance of both the RISBFNN and SDR approaches improve as the number of antennas increases.

The performance of SDR, RISBFNN and random phase is shown in Table \ref{PerformanceRatio}. As we can see, the proposed RISBFNN significantly outperforms random phase, and can achieve decent performance as the SDR approach in various system configurations.

\begin{table}[htbp]
\begin{tabular}{|c|c|c|c|}
\hline
\diagbox{Parameters}{Algorithm} & SDR & RISBFNN & Random\\ \hline
$M=2,N=8$ & 1.6683 & 96.26\% & 65.38\% \\ \hline
$M=2,N=16$ & 2.2814 & 93.90\% & 51.07\% \\ \hline
$M=2,N=32$ & 3.4803 & 92.64\% & 38.77\% \\ \hline
$M=4,N=8$ & 2.2583 & 94.81\% & 76.61\% \\ \hline
$M=4,N=16$ & 2.8477 & 90.65\% & 64.64\% \\ \hline
$M=4,N=32$ & 3.9087 & 88.12\% & 53.18\% \\ \hline
\end{tabular}
\centering
\caption{For the SDR approach, we do 100 randomizations and select the solution with best performance. For the random phase, we randomly select a phase between 0 and $2\pi$ for each reflecting element. The performance of RISBFNN and random phase is shown as the percentage of SDR's performance.}
\label{PerformanceRatio}
\end{table}

\subsection{Computation Complexity}
The complexity of SDR approach is $\mathcal{O}(N^{6.5})$\cite{complexity}, while the complexity of RISBFNN is only $\mathcal{O}((64M+804)N^2)$. The average running time consumed by both algorithms under various system setups are compared in Table \ref{time}. For a fair comparison, both algorithms are executed on the same Intel i7-8700 CPU. P4 is solved by the popular convex optimization solver CVX \cite{CVX}. As we can see, the RISBFNN runs thousands of times faster than the SDR approach.

\begin{table}[htbp]
\begin{tabular}{|c|c|c|}
\hline
\diagbox{Parameters}{Algorithm} & RISBFNN (ms) & SDR (ms)\\ \hline
$M=2,N=16$ & 0.0399 & 199.2\\ \hline
$M=4,N=32$ & 0.0487 & 287.7\\ \hline
$M=8,N=64$ & 0.1167 & 715.3\\ \hline
\end{tabular}
\centering
\caption{Average time consumption of two algorithms.}
\label{time}
\end{table}

In practice, the computation time of passive beamforming algorithm should not exceed the channel coherence time. In the considered indoor scenario, assuming the maximal moving speed being $v_{max}=1.5$ m/s and the frequency being $f_c=2.6$ GHz, the channel coherence time can be computed as $T_c=13.77$ ms. As such, the SDR approach is inapplicable, which makes RISBFNN a promising candidate for practical implementation.

\section{CONCLUSION}
\label{conclusion}
In this paper, we have developed an unsupervised learning based approach for passive beamforming in RIS-assisted communication systems. Through extensive simulations, it has been demonstrated that the proposed approach is capable of performing real-time phase shift configuration while maintaining decent rate performance. In the future, we will consider to extend the proposed framework to more challenging multiuser case. Besides, the use of advanced learning approaches like multi-modal DL\cite{multimodal} to improve system performance is also a research direction worth investigating.

\end{document}